\documentclass[RNAAS]{aastex63}

\shorttitle{}
\shortauthors{Massari, Davide}
\usepackage{hyperref}

\begin{document}

\title{Origin of the system of globular clusters in the Milky Way - {\it Gaia} eDR3 edition}

\author[0000-0001-8892-4301]{Davide Massari}
\affiliation{INAF - Osservatorio di Astrofisica e Scienza dello Spazio di Bologna, Via Gobetti 93/3, I-40129 Bologna, Italy}

\begin{abstract}
In this research note I update the associations between globular clusters and their putative galaxy progenitors determined in \cite{massari19}, based on the kinematic measurements from the {\it Gaia} early data release 3 (eDR3, \citealt{brown21}). The table with the associations is available at {\tt \url{https://www.oas.inaf.it/en/research/m2-en/carma-en/}}, and will be kept up-to-date whenever improved data become available. The same table will also provide updated age estimates from the CARMA project \citep[see][]{massari23}. Please cite this Note alongside {\it Massari, Koppelman \& Helmi 2019, eDR3 edition}.
\end{abstract}

\section{Introductory considerations}

 The sample analyzed here includes the 170 globular clusters (GCs) investigated by \cite{vb2021}.
 In the following, I list the most important aspects of the current analysis.
 
 \begin{itemize}
 \item The orbital parameters employed to repeat the analysis of \citealt{massari19} (hereafter MKH19), namely energy ($E$), vertical angular momentum ($L_{\rm Z}$), perpendicular angular momentum ($L_{\rm perp}$), maximum vertical height ($Z_{\rm max}$), apocenter ({\sf apo}) and circularity ({\sf circ}) are computed by adopting the gravitational potential of \cite{mcmillan17} and using the {\sf AGAMA} code by \cite{vasiliev19}.
 The input 6-D phase space parameters are chosen as follows. Mean positions and proper motions are taken from \cite{vb2021}. Line-of-sight velocities come from \cite{baumgardt19}. The associations are determined on the basis of two different sets of distances. One is the mean distance in \cite{bv2021}, and the other is by \cite{harris96}, 2010 edition. 
 \item The list of GC progenitors studied in MKH19 has been updated with a few additions to reduce the uncertainty about the associations of the so-called high-energy group. The additional progenitors are Elqui (\citealt{shipp19}) and Cetus (\citealt{newberg09}). Their dynamical properties are taken from their discovery papers and from \cite{bonaca21}. 
 I decided not to add the plethora of new candidate mergers found in the halo \citep[e.g.,][]{malhan21, dodd23} because of the uncertainty in the accurate definition of their properties. Moreover, I decided not to split the Sequoia region of the Integrals of Motion (IoM) space into its several substructures \cite[see e.g,][]{dodd23, ceccarelli24}, although it is not incorrect to interpret the Sequoia GCs as possibly associated with Antaeus/ED-3.
 Finally, I maintain the nomenclature "low-energy" GCs used in MKH19, regardless of the debate involving their relation to the Kraken (\citealt{kruijssen20}) and Heracles (\citealt{horta21}) structures.
 \item In general, I maintain the $L_{\rm Z}$, $L_{\rm perp}$ and E boundaries as close as possible to those used in MKH19. Sometimes, they have been adjusted to take into account more recent literature.
 \end{itemize}

\section{Assignment of clusters}

In the following I discuss the results obtained when using the two different sets of distances. I will refer to those from \cite{bv2021} as {\it BV21}, and to those from \cite{harris96} as {\it Ha96}. All the associations are summarized in Table \ref{tab1}. 

\subsection{In situ clusters} 

Bulge clusters (M-B) are selected as those having ${\sf apo}<3.8$ kpc. This limit is set by the inclusion of NGC6171, that according to the age-metallicity criteria in MKH19 is {\it in-situ}, and would result as accreted with a lower apocenter cut.
As in MKH19, Disk clusters (M-D) are selected as those having ${\sf circ}>0.5$ and $Z_{\rm max}<5$ kpc.
By adopting BV21 distances, the resulting number of {\it in-situ} clusters is $74$, of which $44$ are M-B and $30$ are M-D. Compared to MKH19, there are $12$ more {\it in-situ} clusters overall. Among these, $8$ were not in the list of MKH19 GCs.
When using Ha96 distances, the number of {\it in-situ} clusters becomes $72$. Of these, $43$ objects are M-B, and $29$ are M-D.

\subsection{Accreted clusters} 

As in MKH19, GCs are associated with a progenitor based on the comparison between the location of the GCs and that of the stellar debris of known merger events in the IoM space. \\

{\it Sagittarius} (Sag)\\
I refine the MKH19 selection in $L_{\rm Z}$, $E$ and $L_{\rm perp}$ based on the associations made in \cite{bellazzini20}, who exclude NGC5824. Adding a cut in $E$ to the old selection removes this cluster. The final selection is within $3700<L_{\rm perp}<6900$ km/s~kpc, $500<L_{\rm Z}<4500$ km/s~kpc, $-100000<E<-65000$ km$^2$/s$^2$ and includes 9 GCs. 
Compared to the associations in MKH19, besides the loss of NGC5824, two clusters that were not analyzed are added, namely Mu\~{n}oz~1 and Laevens~3.\\

{\it Helmi stream} (H99)\\
The selection is within $600<L_{\rm Z}<2400$  km/s~kpc, $600<L_{\rm perp}<3200$ km/s~kpc,  $E<-100000$ km$^2$/s$^2$, and includes 10 clusters, plus NGC7078 as a candidate member in BV21, only. Among these ten, NGC5053, NGC5024 and Pal5 have been listed as possibly associated with Wukong/LMS-1 (which we do not consider here) by \cite{malhan21}. 
E3 and Rup106 remain listed as uncertain members of H99 (see MKH19). \\

{\it Gaia-Enceladus} (G-E)\\
Clusters are associated with G-E if they have $-640<L_{\rm Z}<620$ km/s~kpc, $-186000<E<-90000$ km$^2$/s$^2$. By using BV21 distances, 26 G-E clusters are found, plus 9 tentative members. Six out of these 9 were already listed as uncertain members in MKH19. Given that the eDR3 data do not unambiguously clarify their association, I maintain the same definition. Among them, NGC5634 and NGC5904 are located in a region of the IoM space that makes them possibly consistent with both G-E and H99.
When using Ha96 distances, the number of G-E GCs becomes 28, plus 8 tentative members.\\

{\it Sequoia} (Seq)\\
Sequoia clusters are those within $-3700<L_{\rm Z}<-640$ km/s~kpc, $-150000<E<-80000$ km$^2$/s$^2$. Their number is 9, but one of these (namely NGC5694) is significantly older than the others, and is therefore excluded based on the age-metallicity considerations of MKH19. The associations are the same as in MKH19, with the addition of AM~4. When using Ha96 distances, the number of associations drops to 7.\\

{\it Low-energy group} (low-E)\\
The selection criteria are unchanged compared to MKH19, with all of the still unassociated clusters having $E<-186000$ km$^2$/s$^2$ being considered as members. When adopting BV21 distances, $20$ clusters are included (plus 1 tentative member, namely UKS1). The adoption of Ha96 distances increases the number of members to $24$. \\

{\it Elqui}\\
Three of the clusters that were unassociated in MKH19 are clustered together at the location of the Elqui stream, at $-1800<L_{\rm Z}<500$ km/s~kpc, $-80000<E<-50000$ km$^2$/s$^2$, $L_{\rm perp}>5000$ km/s~kpc, and are therefore associated with this new progenitor.\\

{\it Cetus}\\
Two clusters are included in the wide prograde area covered by the Cetus stream at $2000<L_{\rm Z}<5000$ km/s~kpc, $E>-90000$ km$^2$/s$^2$, $L_{\rm perp}>8000$ km/s~kpc, namely Ko1 and Pal3.
NGC5824 has been claimed as the possible nuclear star cluster of Cetus by \cite{chang20}, but according to our analysis its dynamics does not fit this definition, neither in $E$ nor in $L_{\rm perp}$.\\

{\it High-energy} (H-E)\\
When using BV21 distances, the remaining unassociated clusters are 11, of which 4 were not analyzed in MKH19. When using Ha96 distances, the number of unassociated clusters is 14. 
These high-energy clusters are likely {\it ex-situ} objects, whose unknown progenitors are low-mass and recently accreted.\\

\begin{table*}
\centering
\caption{\small List of Milky Way GCs and associated progenitors. Clusters in boldface share the same associations in the two cases analyzed.}\label{tab1}
\footnotesize
\begin{tabular}{ccc|ccc|ccc}
\hline
GC & BV2021 & Ha96 & GC &  BV2021 & Ha96 & GC & BV2021 & Ha96 \\
\hline
   {\bf NGC~104}    &  M-D       &   M-D     &  {\bf NGC~6144}     &   low-E      &   low-E   &  {\bf Djorg~2}       &   M-B       &  M-B    \\
   {\bf NGC~288}    &  G-E       &   G-E     &  {\bf NGC~6139}     &   M-B       &   M-B    &  {\it NGC~6517}      &   M-B       &  low-E   \\
   {\bf NGC~362}    &  G-E       &   G-E     &  {\bf Terzan~3}     &   M-B       &   M-B    &  {\it Terzan~10}     &   G-E       &  low-E   \\
   {\bf Whiting~1}  &  Sag       &   Sag     &  {\bf NGC~6171}     &   M-B       &   M-B    &  {\bf NGC~6522}      &   M-B       &  M-B    \\
   {\bf NGC~1261}   &  G-E       &   G-E     &  {\bf 1636-283}     &   M-B       &   M-B    &  {\bf NGC~6535}      &   low-E      &  low-E   \\
   {\bf Pal~1}      &  M-D       &  M-D      &  {\bf NGC~6205}     &   G-E       &   G-E    &  {\bf NGC~6528}      &   M-B       &  M-B    \\
   {\bf AM~1}       &  Elqui     &  Elqui    &  {\bf NGC~6229}     &   G-E       &   G-E    &  {\bf NGC~6539}      &   M-B       &  M-B    \\
   {\bf Eridanus}   &  Elqui     &  Elqui    &  {\bf NGC~6218}     &   M-D       &   M-D    &  {\bf NGC~6540}      &   M-B       &  M-B    \\
   {\bf Pal~2}      &  G-E?      &   G-E?    &  {\bf FSR~1735}     &   low-E      &   low-E   &  {\bf NGC~6544}      &   low-E      &  low-E   \\
   {\bf NGC~1851}   &  G-E       &   G-E     &  {\bf NGC~6235}     &   H99       &    H99   &  {\bf NGC~6541}      &   low-E      &  low-E   \\
   {\bf NGC~1904}   &  G-E       &   G-E     &  {\bf NGC~6254}     &   low-E      &   low-E   &  {\bf ESO-SC06}      &   G-E       &  G-E    \\
   {\it NGC~2298}   &  G-E       &   Seq     &  {\it NGC~6256}     &   M-B       &   low-E   &  {\it NGC~6553}      &   M-D       &  M-B    \\
   {\bf NGC~2419}   &  Sag       &   Sag     &  {\bf Pal~15}       &   G-E?      &   G-E?   &  {\bf NGC~6558}      &   M-B       &  M-B    \\
   {\bf Ko~2}       &  H-E       &   H-E     &  {\bf NGC~6266}     &   M-B       &   M-B    &  {\bf IC~1276}       &   M-D       &  M-D    \\
   {\bf Pyxis}      &  Elqui     &  Elqui    &  {\bf NGC~6273}     &    low-E     &   low-E   &  {\bf Terzan~12}     &   M-D       &  M-D    \\
   {\bf NGC~2808}   &  G-E       &   G-E     &  {\it NGC~6284}     &   low-E      &   G-E    &  {\bf NGC~6569}      &   M-B       &  M-B    \\
   {\bf E~3}        &  H99?      &   H99?    &  {\bf NGC~6287}     &    low-E     &   low-E   &  {\bf BH~261}        &   M-B       &  M-B    \\
   {\bf Pal~3}      &  Cetus     &  Cetus    &  {\bf NGC~6293}     &   M-B       &   M-B    &  {\bf NGC~6584}      &   H99/G-E   &  H99/G-E\\
   {\bf NGC~3201}   &  Seq/G-E   &  Seq/G-E  &  {\bf NGC~6304}     &   M-B       &   M-B    &  {\bf Mercer~5}      &   M-D       &  M-D    \\
   {\bf ESO~93-8}   &  M-D       &  M-D      &  {\it NGC~6316}     &   low-E      &   M-B    &  {\bf NGC~6624}      &   M-B       &  M-B    \\
   {\bf Pal~4}      &  H-E       &  H-E      &  {\bf NGC~6341}     &   G-E       &   G-E    &  {\bf NGC~6626}      &   M-B       &  M-B    \\
   {\bf Crater}     &  H-E       &  H-E      &  {\bf NGC~6325}     &   M-B       &   M-B    &  {\bf NGC~6638}      &   M-B       &  M-B    \\
   {\bf Bliss~1}    &  H-E       &  H-E      &  {\bf NGC~6333}     &   G-E       &   G-E    &  {\bf NGC~6637}      &   M-B       &  M-B    \\
   {\bf Ko~1}       &  Cetus     &  Cetus    &  {\bf NGC~6342}     &   M-B       &   M-B    &  {\bf NGC~6642}      &   M-B       &  M-B    \\
   {\bf NGC~4147}   &  G-E       &   G-E     &  {\bf NGC~6356}     &   M-D       &   M-D    &  {\bf NGC~6652}      &   M-B       &  M-B    \\
   {\bf NGC~4372}   &  M-D       &   M-D     &  {\bf NGC~6355}     &   M-B       &   M-B    &  {\bf NGC~6656}      &  M-D        &  M-D    \\
   {\bf Rup~106}    &  H99?      &   H99?    &  {\bf NGC~6352}     &   M-D       &   M-D    &  {\bf Pal~8}         &   M-D       &  M-D    \\
   {\bf NGC~4590}   &  H99       &   H99     &  {\bf IC~1257}      &   G-E       &   G-E    &  {\bf NGC~6681}      &   low-E      &  low-E   \\
   {\bf BH~140}     &  M-D       &   M-D     &  {\bf Terzan~2}     &   M-B       &   M-B    &  {\bf Ryu~879}       &  G-E        &  G-E    \\
   {\bf NGC~4833}   &  G-E       &   G-E     &  {\bf NGC~6366}     &   M-D       &   M-D    &  {\bf NGC~6712}      &   low-E      &  low-E   \\
   {\bf NGC~5024}   &  H99    &  H99   &  {\bf Terzan~4}     &    M-B      &   M-B    &  {\bf NGC~6715}      &   Sag       &  Sag    \\
   {\bf NGC~5053}   &  H99    &  H99   &  {\bf HP~1}         &    M-B      &   M-B    &  {\bf NGC~6717}      &   M-B       &  M-B    \\
   {\bf Kim~3}      &  H-E       &  H-E      &  {\bf FSR~1758}     &    Seq      &   Seq    &  {\bf NGC~6723}      &   M-B       &  M-B    \\
   {\bf NGC~5139}   & G-E/Seq    & G-E/Seq   &  {\bf NGC~6362}     &    M-D      &   M-D    &  {\bf NGC~6749}      &  M-D        &  M-D    \\
   {\bf NGC~5272}   &  H99       &   H99     &  {\bf Liller~1}     &   M-B       &   M-B    &  {\bf NGC~6752}      &  M-D        &  M-D    \\
   {\bf NGC~5286}   &  G-E       &   G-E     &  {\bf NGC~6380}     &   M-B       &   M-B    &  {\bf NGC~6760}      &  M-D        &  M-D    \\
   {\it AM~4}       &  Seq       &   H-E     &  {\bf Terzan~1}     &    M-B      &   M-B    &  {\bf NGC~6779}      &   G-E       &  G-E    \\
   {\bf NGC~5466}   &  Seq       &   Seq     &  {\bf Ton2}         &   M-B       &   M-B    &  {\bf Terzan~7}      &   Sag       &  Sag    \\
   {\bf NGC~5634}   & G-E/H99   & G-E/H99   &  {\it NGC~6388}     &   low-E      &   M-B    &  {\bf Pal~10}        &   M-D       &  M-D    \\
   {\bf NGC~5694}   &  H-E       &  H-E      &  {\bf NGC~6402}     &   low-E      &   low-E   &  {\bf Arp~2}         &   Sag       &  Sag    \\
   {\bf IC~4499}    &  Seq       &   Seq     &  {\it NGC~6401}     &   M-B       &   low-E   &  {\bf NGC~6809}      &   low-E      &  low-E   \\
   {\bf Mu\~{n}oz~1}    &  Sag       &   Sag     &  {\bf NGC~6397}     &   M-D       &   M-D    &  {\bf Terzan~8}      &   Sag       &  Sag    \\
   {\bf NGC~5824}   &  H-E       &  H-E      &  {\bf VVV~CL002}    &   M-B       &   M-B    &  {\bf Pal~11}        &   M-D       &  M-D    \\
   {\bf Pal~5}      &  H99?      &   H99?    &  {\it Pal~6}        &   M-B       &   low-E   &  {\bf NGC~6838}      &   M-D       &  M-D    \\
   {\bf NGC~5897}   &  G-E       &   G-E     &  {\bf NGC~6426}     &   H99      &   H99   &  {\bf NGC~6864}      &   G-E       &  G-E    \\
   {\bf NGC~5904}   &G-E/H99     &G-E/H99    &  {\it Djorg~1}      &   G-E       &   H-E    &  {\bf NGC~6934}      &  H-E        & H-E     \\
   {\bf NGC~5927}   &  M-D       &   M-D     &  {\bf Terzan~5}     &   M-B       &   M-B    &  {\bf NGC~6981}      &  G-E        &  G-E    \\
   {\bf NGC~5946}   &  low-E      &   low-E    &  {\bf NGC~6440}     &   M-B       &   M-B    &  {\it NGC~7006}      &  Seq        & H-E     \\
   {\bf BH~176}     &  M-D       &  M-D      &  {\it NGC~6441}     &   M-D       &   low-E   &  {\bf Laevens~3}     &   Sag       &  Sag    \\
   {\bf NGC~5986}   &  low-E      &   low-E    &  {\bf Terzan~6}     &   M-B       &   M-B    &  {\bf Segue~3}       &   H-E       & H-E     \\
   {\bf FSR~1716}   &  M-D       &   M-D     &  {\it NGC~6453}     &   M-B       &   low-E   &  {\it NGC~7078}      &   H99?      &  M-D    \\
   {\bf Pal~14}     &  H-E       &  H-E      &  {\it UKS~1}        &   G-E/low-E  &   M-B    &  {\bf NGC~7089}      &   G-E       &  G-E    \\
   {\bf Lynga~7}    &  M-D       &   M-D     &  {\bf VVV~CL001}    &   low-E      &   low-E   &  {\bf NGC~7099}      &   G-E       &  G-E    \\
   {\bf NGC~6093}   &  low-E      &   low-E    &  {\bf Gran~1}       &   M-B       &   M-B    &  {\bf Pal~12}        &   Sag       &  Sag    \\
   {\bf Ryu~059}    &  H-E       &   H-E     &  {\bf Pfleiderer2}  &   M-D       &   M-D    &  {\bf Pal~13}        &   Seq       &  Seq    \\
   {\bf NGC~6121}   &  low-E      &   low-E    &  {\bf NGC~6496}     &   M-D       &   M-D    &  {\bf NGC~7492}      &   G-E       &  G-E    \\
   {\bf NGC~6101}   &  Seq/G-E   &  Seq/G-E  &  {\bf Terzan~9}     &   M-B       &   M-B    &      -               &             &         \\
\hline                                
\end{tabular}
\end{table*}

\bibliographystyle{aasjournal}

\end{document}